\documentclass[useAMS,usenatbib]{mn2e}

\usepackage{times,graphicx,amssymb,natbib}

\newcommand{\alphacen}{$\alpha$ Cen }
\newcommand{\degrees}{^{\circ}}
\newcommand{\msol}{M_{\rm \odot}}
\newcommand{\mjup}{M_{\rm Jup}}

\title[The Habitable Zone around $\alpha$ Centauri B]{Oscillations in
  the Habitable Zone around $\alpha$ Centauri B}
\author[D. Forgan]{Duncan Forgan $^{1}$\thanks{E-mail: dhf@roe.ac.uk}
  \\ $^{1}$Scottish Universities Physics Alliance (SUPA), Institute
  for Astronomy, University of Edinburgh, Blackford Hill, Edinburgh,
  EH9 3HJ, Scotland, UK\\}
\begin{document}

\date{Accepted}

\pagerange{\pageref{firstpage}--\pageref{lastpage}} \pubyear{}

\maketitle

\label{firstpage}

\begin{abstract}

\noindent The \alphacen AB system is an attractive one for radial
velocity observations to detect potential exoplanets.  The high
metallicity of both \alphacen A and B suggest that they could have
possessed circumstellar discs capable of forming planets. As the
closest star system to the Sun, with well over a century of accurate
astrometric measurements (and \alphacen B exhibiting low chromospheric
activity) high precision surveys of \alphacen B's potential
exoplanetary system are possible with relatively cheap
instrumentation.  Authors studying habitability in this system
typically adopt habitable zones (HZs) based on global radiative
balance models that neglect the radiative perturbations of \alphacen
A.

We investigate the habitability of planets around \alphacen B using 1D
latitudinal energy balance models (LEBMs), which fully incorporate the
presence of \alphacen A as a means of astronomically forcing
terrestrial planet climates.  We find that the extent of the HZ is
relatively unchanged by the presence of \alphacen A, but there are
variations in fractional habitability for planets orbiting at the
boundaries of the zone due to \alphacen A, even in the case of zero
eccentricity.  Temperature oscillations of a few K can be observed at
all planetary orbits, the strength of which varies with the planet's
ocean fraction and obliquity.

\end{abstract}


\section{Introduction}

\noindent There are currently 755 known extrasolar planets, spanning a
large parameter space in orbital elements and in planetary
properties\footnote{http://exoplanet.eu/catalog.php}.  The question of
planet formation in binary systems is an important one, as a large
fraction of solar type stars are born in binary or multiple star
systems \citep{Duquennoy1991}.  The majority of these systems are wide
S-type binaries, where the secondary orbits at distances of over 100
au from the planetary system.  At the time of writing, around 15\% of
planets have been detected in binary star systems
\citet{Desidera2007}. In S-type systems, planet formation modes should
be similar to those in single star systems, with the secondary's
influence being roughly negligible (unless its orbit is sufficiently
eccentric).

However, there are a handful of systems with detected planets where
the secondary orbits at much closer separations ($\sim 20$ au) -
$\gamma$ Cephei b \citep{Hatzes2003}, HD41004b \citep{Zucker2004},
GJ86b \citep{Queloz2000}, etc.  The $\alpha$ Centauri system has
similar orbital architecture, although as yet it does not host
detections of exoplanets.

Despite this, the \alphacen system is an attractive one for studying
possible planet formation in binary systems.  At a distance of 1.33
pc, it is the nearest star system to our Sun.  It is composed of a
hierarchical triple system, with the central binary system (\alphacen
AB) orbited by the M dwarf Proxima Cen at sufficiently large distance
to be negligible \citep{Wertheimer2006}.  \alphacen A is type G2V with
mass $M_A = 1.105 \pm 0.007 \msol$, and \alphacen B is type K1V with
mass $M_B = 0.934 \pm 0.007 \msol$.  It is particularly amenable to
relatively cheap radial velocity (RV) campaigns \citep{Guedes2008},
especially if the goal is to detect Earth-mass planets within the
habitable zone of a solar-type star.

As \citet{Guedes2008} describe, there are several reasons why this is the case:
firstly, both \alphacen A and B are high metallicity stars, which
would promote the existence of circumstellar discs with a high
fraction of solid materials at early times \citep{Wyatt_Z}, as well as
deeper spectral lines for improved RV precision.  \alphacen B is
particularly quiet in terms of chromospheric activity and acoustic
p-wave oscillations, with a relatively strong potential RV signal due
to its lower mass.  The binary is inclined by only 11 degrees with
respect to the line of sight, i.e. its inclination angle from face-on
is $i=79^{\circ}$.  This ensures that $\sin i \approx 1$, and
that the recovered planet mass from RV observations will be close to
the true mass, provided that the planets form and remain in the same plane.
Finally, its position in the sky ($-60 \degrees$ declination) affords
astronomers in the southern hemisphere the opportunity to observe the
system for most of the year.  If \alphacen B does host terrestrial
planets, they should be readily detectable even with a 1 metre
telescope with high-resolution spectrograph instrumentation.

Several numerical studies have shown that both stars are capable of
forming terrestrial planets despite the perturbing influence of the
binary companion, which appears to play a role analogous to the gas
giants within our solar system.  The planetesimal discs appear to be
stable within 3 AU of their parent stars, provided the inclination of
the disc relative to the binary plane is less than 60 degrees
\citep{Wiegert1997,Quintana2002,Quintana2007}.  However, other studies
have shown that the later stages of accretion to produce lunar mass
objects is reduced in efficiency due to orbital rephasing by the
binary companion.  This inhibits collisional growth around \alphacen A
to regions within 0.75 au \citep{Thebault2008}, and within 0.5 au of
\alphacen B \citep{Thebault2009}.

It therefore appears to be the case that gas giant formation is
suppressed relative to our solar system \citep{Xie2010}, a prediction
consistent with the absence of detections from previous radial
velocity surveys, which suggest upper limits between 0.5 and 3
$\mjup$.  If the disc can produce a sufficient quota of lunar mass
bodies, \citet{Guedes2008} show that Earth mass planets can form in
the habitable zone of \alphacen B, with maximum eccentricities of
around $e_P=0.3$ (this result is also corroborated by Xie et al 2010).

Once formed, Hamiltonian analysis indicates terrestrial planets
orbiting in \alphacen B's habitable zone appear to be dynamically
stable under perturbations from \alphacen A, provided that $e_p<0.3$
and the inclination of the planet's orbit relative to the binary
plane, $i_p< 35 \degrees$ \citep{Michtchenko2009}. Equally, these
authors also show that planets with inclinations larger than this
value are expected to experience strong instability due to the
Lidov-Kozai resonance resulting in eccentricity-inclination coupling.

These studies have generally assumed a habitable zone (HZ) around
\alphacen B in the semi major axis range $0.5 < a < 0.9$ au.  This
range is based on calculations by \citet{Kasting_et_al_93}, who
calculate the HZ using a global radiative balance (GRBM) model,
assuming Earth mass planets with similar atmospheric composition
(N$_2$/H$_2$O/CO$_2$).  The inner edge of the HZ is governed by loss
of water via photolysis and subsequent hydrogen escape, and the outer
edge is determined by formation of CO$_2$ clouds, which increase the
planet's albedo.  In defining the habitable zone in this fashion, the
perturbing influence of \alphacen A is neglected.  Given the semimajor
axis of the orbit, this would appear to be an appropriate
approximation.  If main sequence relations for the luminosity of each
object are assumed, the insolation experienced by planets in the
habitable zone of \alphacen B due to \alphacen A would be no more than
a few percent of the total insolation of the \alphacen AB system at
the binary's periastron, and around one tenth of a percent at
apastron.  This insolation can be diminished further by eclipses of
\alphacen A by \alphacen B, the duration of which is estimated to be
of order a few Earth days.

However, we should also consider the results of more complex 1D
latitudinal energy balance models (LEBMs) such as those described by
\citet{Williams1997a} and subsequently
\citet{Spiegel_et_al_08,Spiegel2009,Dressing2010,Spiegel2010}.  Rather
than pursuing a simple global balance, the planet's temperature is
allowed to vary as a function of latitude, $\lambda$.  The insolation
of the planet will also be a function of latitude and season, and the
other key properties (infrared cooling rate and albedo) are also
temperature dependent.

A planet in global radiative balance is not in general in
local radiative balance, and by extension habitability is not a
discrete concept (i.e. either habitable or uninhabitable), but a
continuous one, where a certain fraction of the planet's surface will
be habitable at any given time.  In the LEBM, the evolution of the planet's
temperature $T(\lambda)$ is described by a diffusion equation made
non-linear by the addition of the heating and cooling terms, as well
as an albedo which makes a rapid transition from low to high as
temperature decreases past the freezing point of water.  As a result,
small changes in the properties of a planet can strongly affect the
resultant climate.  For example, changing the length of day in an
Earth-type model can determine whether the planet can retain liquid
water on the surface, or undergo an albedo feedback reaction which
results in the ``Snowball Earth'' scenario \citep{Spiegel_et_al_08}.

These models have been successful in establishing important aspects of
astronomical forcings on climate, e.g. Milankovitch cycles
\citep{Spiegel2010}, variations due to orbital eccentricity, and
potentially the effects of Kozai resonances or other orbital
instabilities.  It is this propensity for incorporating external
forcing (such as the presence of a binary companion) which lends it
towards studying habitability in systems such as \alphacen AB.  The
sensitivity of the climate to such forcings suggest that the
relatively small perturbation produced by the presence of \alphacen A
may have important consequences for the location of the habitable zone
around \alphacen B.

In this work, we augment the 1D LEBM of \citet{Spiegel_et_al_08} to
include the presence of a binary companion, and perform a parameter
study for planets orbiting \alphacen B.  In particular, we investigate
the currently defined habitable zone, to compare with the GRBM calculations of
\citet{Kasting_et_al_93}. Section \ref{sec:Method} describes our modified
LEBM, and the initial conditions studied in the parameter space
survey.  Section \ref{sec:Results} displays the results of this
study.  In section \ref{sec:Discussion} we discuss the implications of
these results, and summarise the work in section \ref{sec:Conclusions}.

\section{Method }\label{sec:Method}

\subsection{Latitudinal Equilibrium Balance Models}

\noindent At their core, LEBMs solve the following diffusion equation:

\begin{equation} C \frac{\partial T(x,t)}{\partial t} -
  \frac{\partial }{\partial x}\left(D(1-x^2)\frac{\partial
    T(x,t)}{\partial x}\right) = S(1-A(T)) - I(T), \end{equation}

\noindent where $T=T(x,t)$ is the temperature at $x = \sin \lambda$,
and $\lambda$ is the latitude (between $-90\degrees$ and
$90\degrees$).  This equation is evolved with the boundary condition
$\frac{dT}{dx}=0$ at the poles.  The $(1-x^2)$ term is a geometric
factor, arising from solving the diffusion equation in spherical
geometry.

$C$ is the effective heat capacity of the atmosphere, $D$
is a diffusion coefficient that determines the efficiency of heat
redistribution across latitudes, $S$ is the insolation flux, $I$ is
the IR cooling and $A$ is the albedo.  In the above equation, $C$,$S$,$I$
and $A$ are functions of $x$ (either explicitly, as $S$ is, or
implicitly through $T(x)$).   

$D$ is a constant, defined such that a planet at 1 au
around a star of $1 \msol$, with diurnal period of 1 day will
reproduce the average temperature profile measured on Earth.  Planets
that rapidly rotate experience inhibited latitudinal heat transport,
due to Coriolis effects (see \citealt{Farrell1990}).  In the model, we follow
\citet{Spiegel_et_al_08} by scaling $D$ according to:

\begin{equation} D=5.394 \times 10^2 \left(\frac{\omega_d}{\omega_{d,\oplus}}\right)^{-2},\label{eq:D}\end{equation}

\noindent where $\omega_d$ is the rotational angular velocity of the
planet, and $\omega_{d,\oplus}$ is the rotational angular velocity of
the Earth.   This expression is probably too simple to describe the
full effects of rotation, but in the absence of a first-principle
theory to describe that describes reduced transport we must make do.

\noindent In this work, we solve the diffusion equation using a simple
explicit forward time, centre space finite difference algorithm.  A
global timestep was adopted, with constraint

\begin{equation}\delta t < \frac{\left(\Delta x\right)^2C}{2D(1-x^2)}.  \end{equation}

The parameters are diurnally averaged, i.e. a
key assumption of the model is that the planets rotate sufficiently
quickly relative to their orbital period.  We adopt the same
expressions for these parameters as \citet{Spiegel_et_al_08}, who in
turn used the work of \citet{Williams1997a}.

The atmospheric heat capacity depends on what fraction of the planet's
surface is ocean, $f_{ocean}$, what fraction is land $f_{land}=1.0-f_{ocean}$, and
what fraction of the ocean is frozen $f_{ice}$:

\begin{equation} C = f_{land}C_{land} + f_{ocean}\left((1-f_{ice})C_{ocean} + f_{ice} C_{ice}\right). \end{equation}

\noindent The heat capacities of land, ocean and ice covered areas are
\begin{equation} C_{land} = 5.25 \times 10^9 $ erg cm$^{-2}$ K$^{-1}\end{equation}
\begin{equation} C_{ocean} = 40.0C_{land}\end{equation}
\begin{equation} C_{ice} = \left\{
\begin{array}{l l }
9.2C_{land} & \quad \mbox{263 K $< T <$ 273 K} \\
2C_{land} & \quad \mbox{$T<263$ K}. \\
\end{array} \right. \end{equation}

The infrared cooling function is 

\begin{equation} I(T) = \frac{\sigma_{SB}T^4}{1 +0.75
    \tau_{IR}(T)}, \end{equation}

\noindent where the optical depth of the atmosphere 

\begin{equation} \tau_{IR}(T) =
  0.79\left(\frac{T}{273\,\mathrm{K}}\right)^3. \end{equation}

\noindent The albedo function is

\begin{equation} A(T) = 0.525 - 0.245 \tanh \left[\frac{T-268\,
      \mathrm K}{5\, \mathrm K} \right]. \end{equation}

\noindent This produces a rapid shift from low albedo to high albedo
as the temperature drops below the freezing point of water. It
  is this transition that makes the outer habitable zone extremely
  sensitive to changes to various orbital and planetary parameters,
  and makes LEBMs an important tool in studying short-term temporal
  evolution of planetary climates.  The insolation flux $S$ is a
function of both season and latitude.  At any instant, the bolometric
flux received at a given latitude at an orbital distance $r$ is

\begin{equation}S = q_0\cos Z \left(\frac{1 AU}{r}\right)^2,\end{equation}

\noindent where $q_0$ is the bolometric flux received from the star at a
distance of 1 AU, and $Z$ is the zenith angle:

\begin{equation} q_0 = 1.36\times 10^6\left(\frac{M}{\msol}\right)^4
  erg s^{-1} cm^{-2} \end{equation}

\begin{equation} \cos Z = \mu = \sin \lambda \sin \delta + \cos
  \lambda \cos \delta \cos h. \end{equation} 

\noindent We have assumed main sequence scaling for the luminosity
($\msol$ represents one solar mass).  Given that both \alphacen A and
B are similar in mass (and spectral type) to the Sun, this is a
reasonable first approximation, however, we should note the
observational constraints placed by \citet{Thevenin2002}, as we will
in the Discussion. $\delta$ is the solar declination, and
$h$ is the solar hour angle.  The solar declination is calculated from
the obliquity $\delta_0$ as:

\begin{equation} \sin \delta = -\sin \delta_0
  \cos(\phi_p-\phi_{peri}-\phi_a), \end{equation}

\noindent where $\phi_p$ is the current orbital longitude of the
planet, $\phi_{peri}$ is the longitude of periastron, and $\phi_a$ is
the longitude of winter solstice, relative to the longitude of
periastron.  

We must diurnally average the solar flux:

\begin{equation} S = q_0 \bar{\mu}. \end{equation}

\noindent This means we must first integrate $\mu$ over the sunlit part of the
day, i.e. $h=[-H, +H]$, where $H$ is the radian half-day length at a
given latitude.  Multiplying by the factor $H/\pi$ (as $H=\pi$ if
a latitude is illuminated for a full rotation) gives the total diurnal
insolation as

\begin{equation} S = q_0 \left(\frac{H}{\pi}\right) \bar{\mu} = \frac{q_0}{\pi} \left(H
  \sin \lambda \sin \delta + \cos \lambda \cos \delta \sin H\right). \end{equation}

\noindent The radian half day length is calculated as

\begin{equation} \cos H = -\tan \lambda \tan \delta. \end{equation}

\noindent We calculate habitability indices in the same manner as
\citet{Spiegel_et_al_08}.  The habitability function $\eta$  is:

\begin{equation} \eta(\lambda,t) = \left\{
\begin{array}{l l }
1 & \quad \mbox{273 K $< T(\lambda,t) <$ 373 K} \\
0 & \quad \mbox{otherwise}. \\
\end{array} \right. \end{equation}

\noindent We then average this over latitude to calculate
the fraction of habitable surface at any timestep:

\begin{equation} \bar{\eta}(t) =
  \frac{\int_{-\pi/2}^{\pi/2}\eta(\lambda,t)\cos \lambda \,
    d\lambda}{2}. \end{equation}

\noindent We will use this function to classify the planets
we simulate in the following sections.

\subsection{Modifications to include the binary}

\noindent The addition of the second star requires us to repeat the
insolation flux calculation, where we must now re-calculate the
orbital longitude, solar declination and radian half-day length of the
planet relative to the secondary, and use the current the orbital distance
from the planet to the secondary.  

We must also account for the possibility of transits.  Given the orbital
configuration, transits of the secondary (\alphacen A) by the primary
(\alphacen B) should occur frequently in the simulation.  By
calculating the angle between the vector $\mathbf{r}_{21}$ between the
two stars, and the vector $\mathbf{r}_{p1}$ between the planet and the
primary, it can be determined whether a transit is occurring at any
given timestep \footnote{We also assume the stellar radii are governed
  by main sequence relations, see \citet{Prialnik}}.  If the primary
transits the secondary, the secondary flux is set to zero.  

We ensure that transits are resolved in time by adding a second timestep
criterion, which ensures that at the planet's current orbital
velocity, the duration of a transit will not be less than 10
timesteps.  We neglect the distant companion Proxima Centauri.

\subsection{Initial Conditions}

\noindent Unless otherwise stated, Earthlike conditions are assumed.
The diurnal period is equal to the Earth's, and the obliquity is set
to $23.5\degrees$.  $f_{ocean}$ is set to 0.7.

The \alphacen system is set up to have $M_1 = M_B = 0.934 \msol$, $M_2
= M_A = 1.1 \msol$.  The semimajor axis of the orbit is $a=23.4$ au,
and the eccentricity $e=0.5179$.  \alphacen A is placed at apastron at
the beginning of all simulations.

In line with the results of \citet{Guedes2008}, we
investigate planet eccentricities between $e_p=0$ to $e_p=0.3$, and semi
major axes relative to \alphacen B in the conventionally established habitable zone $0.5<a<0.9$.

The simulations begin at the northern winter solstice, which is
assumed to occur at an orbital longitude of $0 \degrees$.  In the case of eccentric
orbits, this is also the longitude of periastron\footnote{Simulations were
carried out where the longitude of periastron was varied.  The
results were not significantly affected, as the eccentricities studied
are relatively low, see \citet{Dressing2010}}.  The planets' initial
temperature was 288 K globally (starting the simulation at
higher temperatures did not significantly affect the result).  The
simulations were carried out for 1000 years (approximately 15 orbits
of \alphacen A around \alphacen B).  The first 200 years of simulation
are ignored, allowing the system to settle down to a quasi-steady
state in all cases. 

\section{Results}\label{sec:Results}

\subsubsection{The Habitable Zone}

\begin{figure}
\begin{center}
\includegraphics[scale = 0.5]{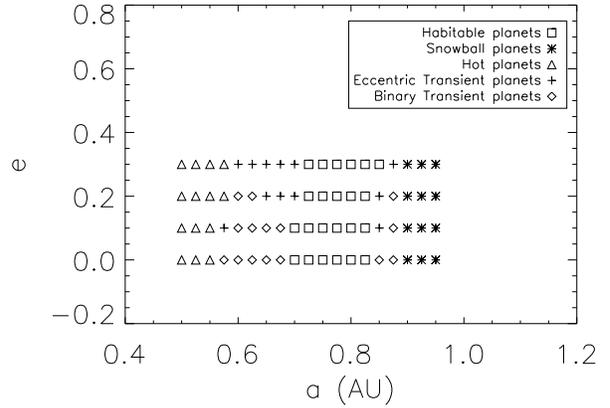}
\caption{The habitable zone around \alphacen B. The x-axis
    shows the semi-major axis of the planet orbiting \alphacen B, the
    y-axis shows the eccentricity of the planet's orbit.  The
  resulting planets are classified into five categories, which are
  described in the text. \label{fig:snowball}}
\end{center}
\end{figure}

\noindent Figure \ref{fig:snowball} shows the end result for all simulations carried
out in this work.  The various end states can be classified thus:

\begin{enumerate}
\item Habitable Planets - these planets are 100\% habitable across the
  entire surface, i.e. $\bar{\eta}(t)=1$ for all $t$.
\item Hot Planets - these planets have temperatures above 373 K across
  all seasons, and are therefore uninhabitable ($\bar{\eta}(t)=0$ for all $t$).
\item Snowball Planets - these planets have undergone a snowball
  transition to a state where the entire planet is frozen, and are
  therefore uninhabitable ($\bar{\eta}(t)=0$ for all $t$).
\item Eccentric Transient Planets - these planets are partially
  habitable ($0<\bar{\eta}(t)<1$), but the habitability fraction
  oscillates according to the period of the planet's orbit around
  \alphacen B.  This oscillation is not present in circular orbits.
\item Binary Transient Planets - these planets are partially habitable
  ($0<\bar{\eta}(t)<1$), but the habitability oscillates with the
  period of \alphacen A. This oscillation is present in both circular
  and eccentric orbits.
\end{enumerate}

\noindent Some systems exhibit both types of transient behaviour.
Where this is the case, we classify systems according to which
oscillation has the largest amplitude. 

If we instead chose to classify by maximum temperature, then most
systems would be regarded as binary transients.  Even planets in the
middle of the habitable zone will experience temperature
fluctuations. The resulting temperature oscillations due to \alphacen
A are quite small - the maximum temperature generally fluctuates by at
most 2K.

\begin{figure}
\begin{center}
\includegraphics[scale = 0.5]{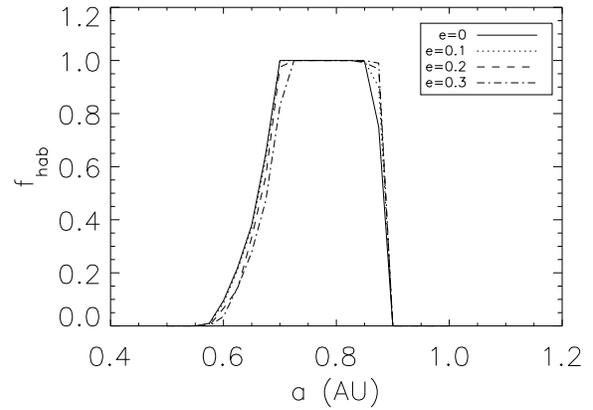}
\caption{Habitability Fraction as a function of planet semi-major
  axis, time averaged in the date range $[300,400]$ years.  Curves
  are plotted for four values of the planet's orbital eccentricity.\label{fig:habvsa}}
\end{center}
\end{figure}

\noindent Figure \ref{fig:habvsa} shows the total fraction of the
planet which is habitable, $\bar{\eta}(t)$, time-averaged over the range
$[300,400]$ years.  The steep gradient at the outer edge of the HZ
indicates the non-linear, highly sensitive nature of the snowball
transition. For single-star models using parameters corresponding to
Earth and the Sun, $\bar{\eta}\approx 0.85$.  Around \alphacen B,
planets on circular orbits will attain similar values at $a_p=0.86$
au \footnote{The same value is achieved at $a_p=0.69$ au, but this
  solution has no ice caps, and an equatorial $T>373$ K}.  This plot
would indicate that the inner HZ would begin nearer to 0.6 than 0.5
au, with the edge of the outer HZ agreeing with previous results. 

\subsection{Habitable, Hot and Snowball Planets}

\noindent Planets in these three categories have stable values of $\eta$ throughout the season.  The
influence of \alphacen A tends to raise the temperature by a few
K at all eccentricities, with this phenomenon insensitive to any
resonances between the longitude of periastron of the star and planet.

Temperature profiles (time averaged over one orbit of the planet) take
essentially the same shape, of the form

\begin{equation} T(\lambda) = A - B \sin^2\lambda \end{equation}

\noindent With the snowball planets taking the lowest value of $B$,
and the hot planets taking the largest values of $B$. 

\subsection{Eccentric Transience}

\begin{figure}
\begin{center}
\includegraphics[scale = 0.5]{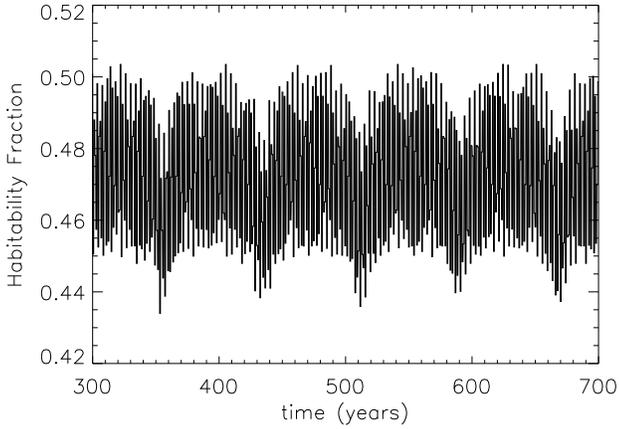}
\caption{Habitability fraction as a function of time for
  the case where $a_p=0.675$, $e_p=0.3$. \label{fig:habfrac_ecctrans}}
\end{center}
\end{figure}

\noindent As was found by \citet{Dressing2010}, eccentric planets are
generally hotter compared to planets in circular orbits at the same
semi-major axis (as received flux scales as $(1-e^2)^{-1/2}$).  As the
eccentricity regime we explore is fairly modest, we do not see planets
entering and leaving the snowball state purely because of high $e_p$.
The planets do not spend long time intervals at apastron (relatively
speaking), and therefore would not be able to freeze except if the
semimajor axis was already sufficiently large.

Figure \ref{fig:habfrac_ecctrans} shows $\bar{\eta}(t)$ in the case
where $a_p=0.675$ and $e_p = 0.3$.  The variations due to the planet's
orbit are clearly visible as rapid fluctuations with amplitude of
approximately 0.02.  While eccentricity is the principal source of
fluctuations here, the influence of \alphacen A is clearly visible
(e.g. at $t\sim 350$ years), reducing the habitable fraction by
$\sim0.01$.  In this example, equatorial latitudes exhibit
temperatures above the boiling point of water - the passage of
\alphacen A increases the thickness of this inhospitable band. 

\subsection{Binary Transience}

\begin{figure*}
\begin{center}$
\begin{array}{cc}
\includegraphics[scale = 0.5]{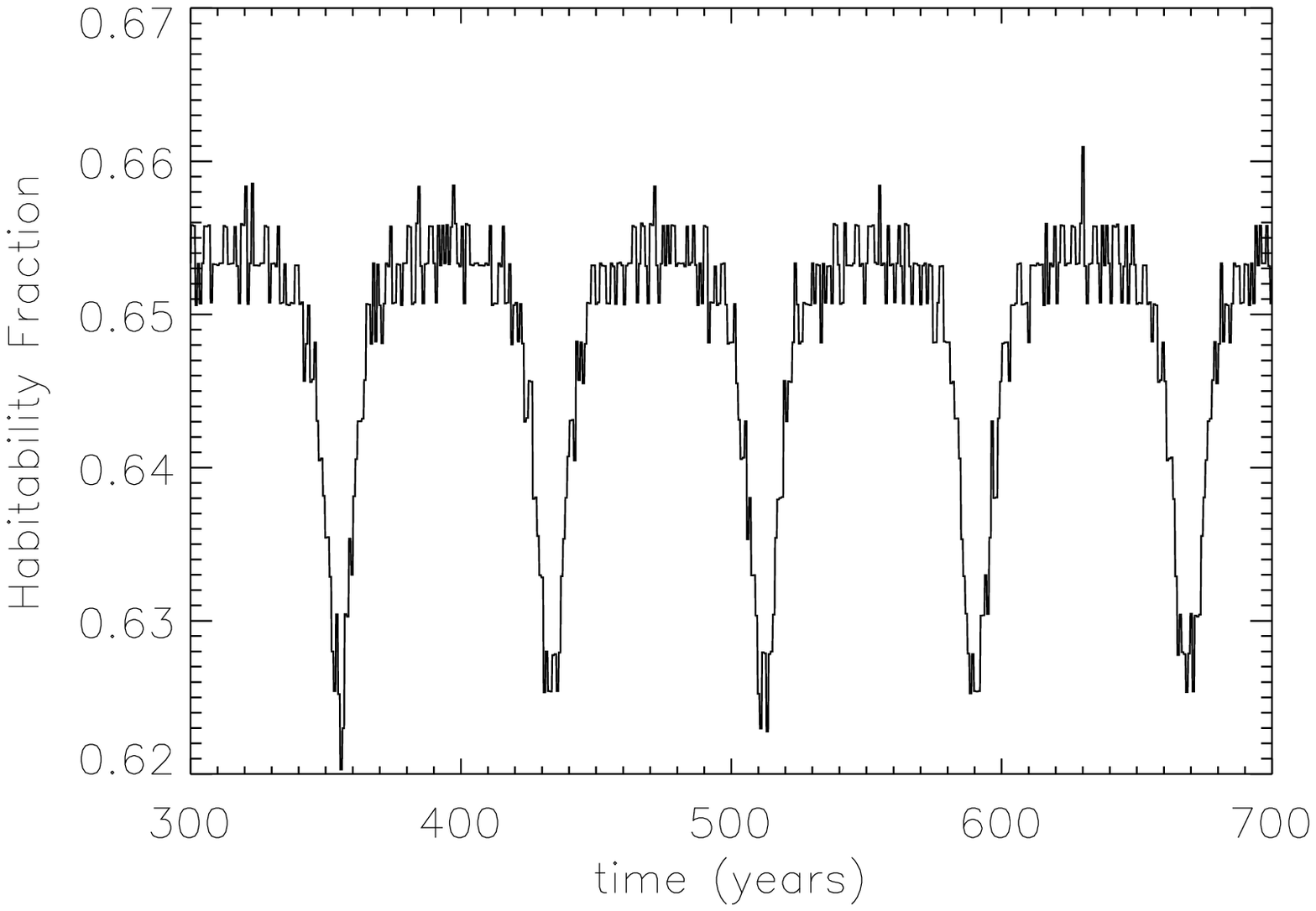} &
\includegraphics[scale=0.5]{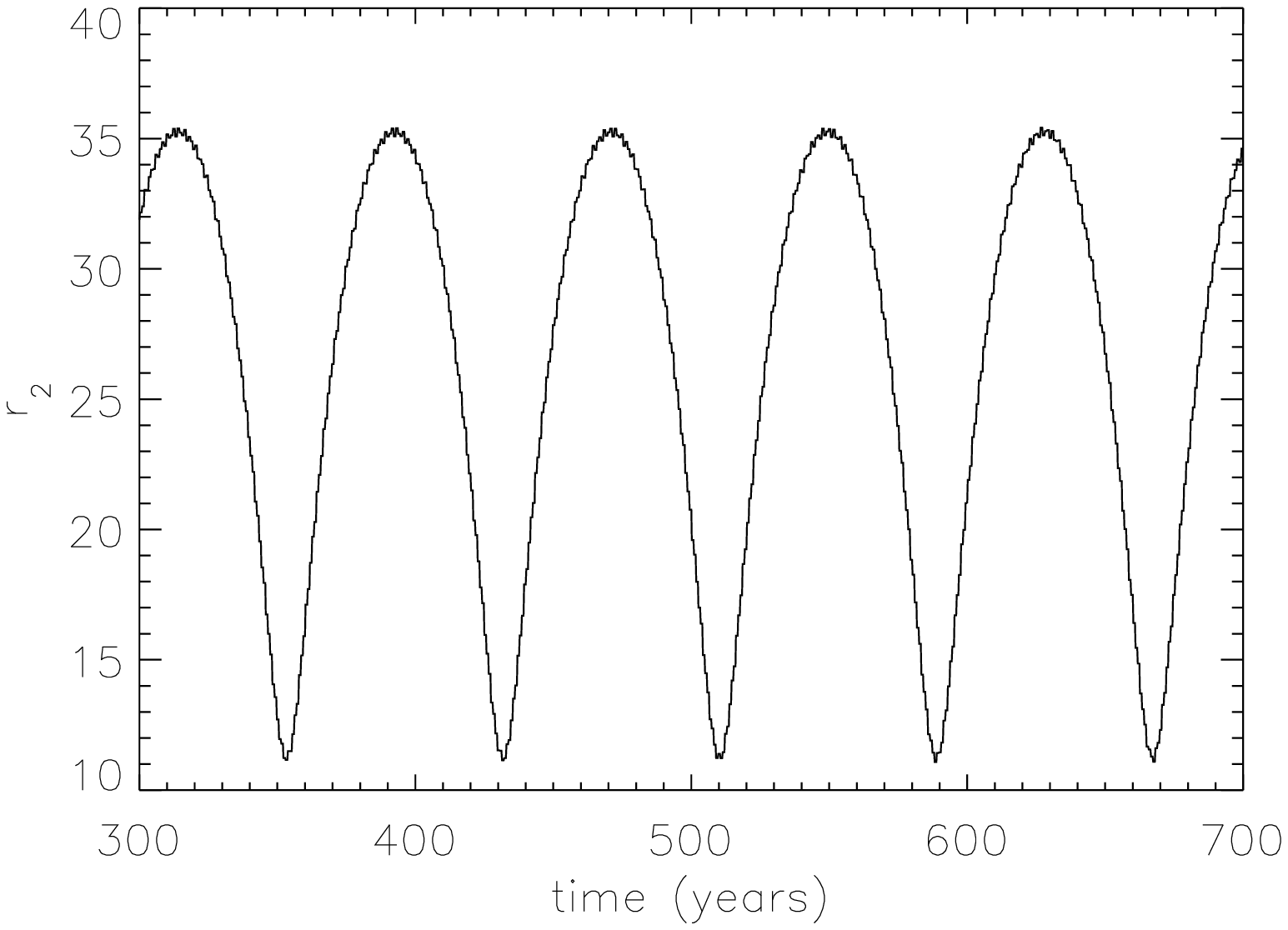}
\end{array}$
\caption{Left:Habitability fraction as a function of time for
  the case where $a_p=0.675$, $e_p=0.0$. Right: distance between the
  secondary and the planet as a function of time for the same planet
  parameters. \label{fig:habfrac_bintrans}}
\end{center}
\end{figure*}

\noindent If we now take the previous example and set $e_p=0$, then we
can see the influence of \alphacen A begin to dominate (left panel of
Figure \ref{fig:habfrac_bintrans}).  The
fluctuations due to eccentricity have effectively disappeared, and
periastron passage of \alphacen A (right panel of Figure
\ref{fig:habfrac_bintrans} induces sharp decreases in
$\bar{\eta}(t)$ of around 0.025.  This clearly shows the habitability of
the planet being significantly affected by \alphacen A.  This is
despite the ratio of mean insolation from the primary and secondary,
$\frac{\bar{S}_2}{\bar{S}_1} \leq 0.01$.

\section{Discussion}\label{sec:Discussion}

\subsection{Dependence on the Secondary's Orbit}

\begin{figure*}
\begin{center}$
\begin{array}{cc}
\includegraphics[scale = 0.45]{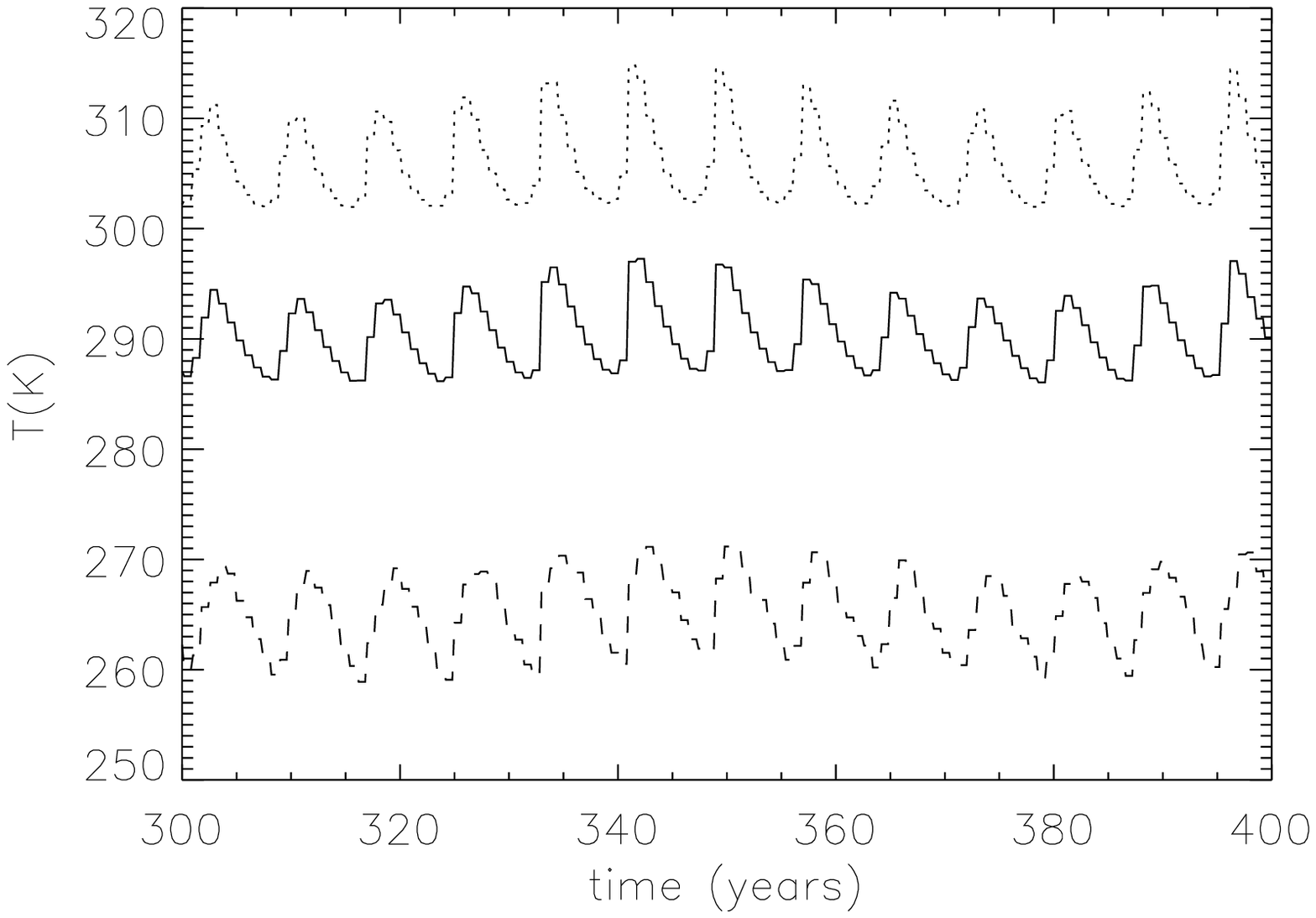} &
\includegraphics[scale = 0.45]{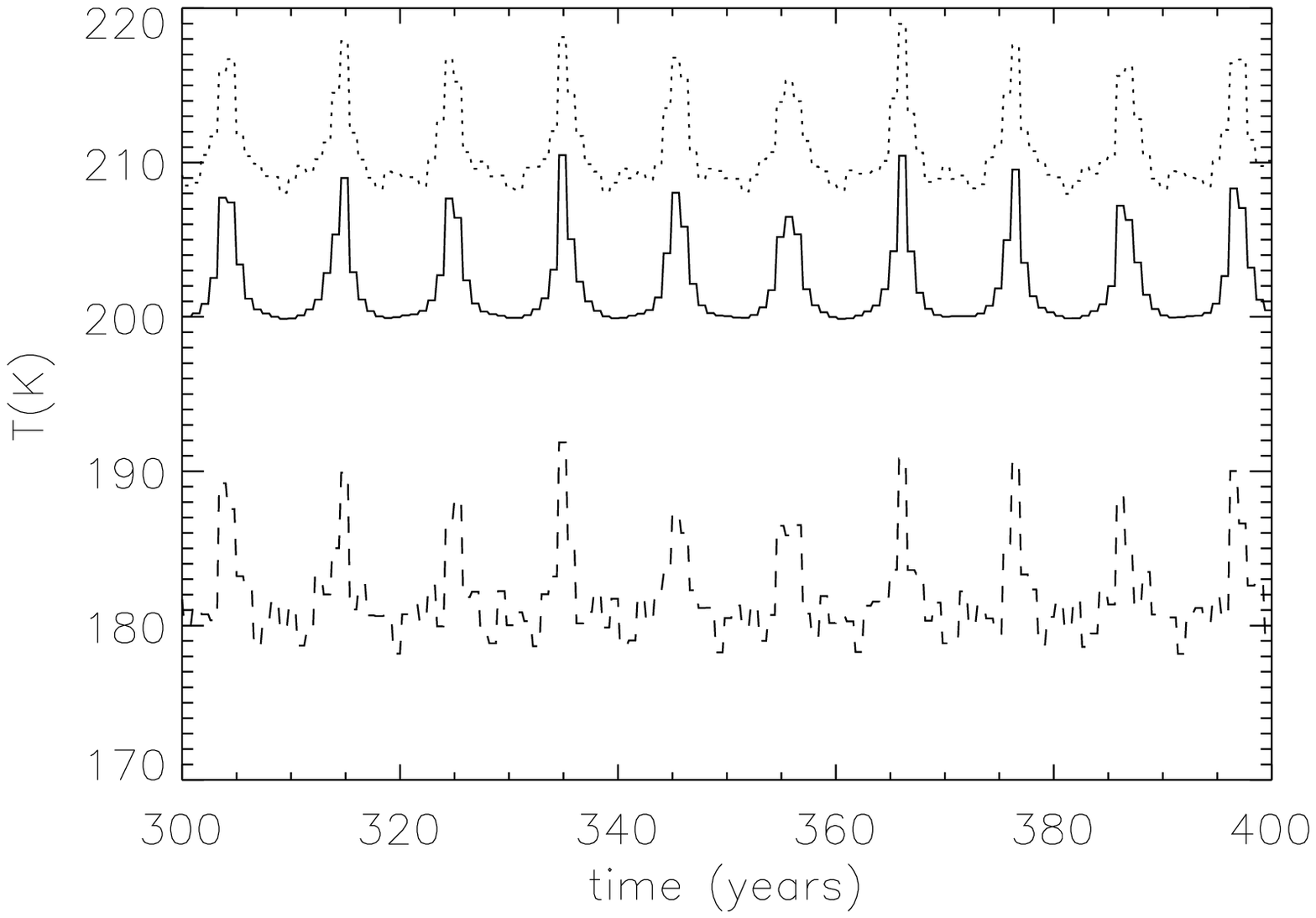} \\
\includegraphics[scale=0.45]{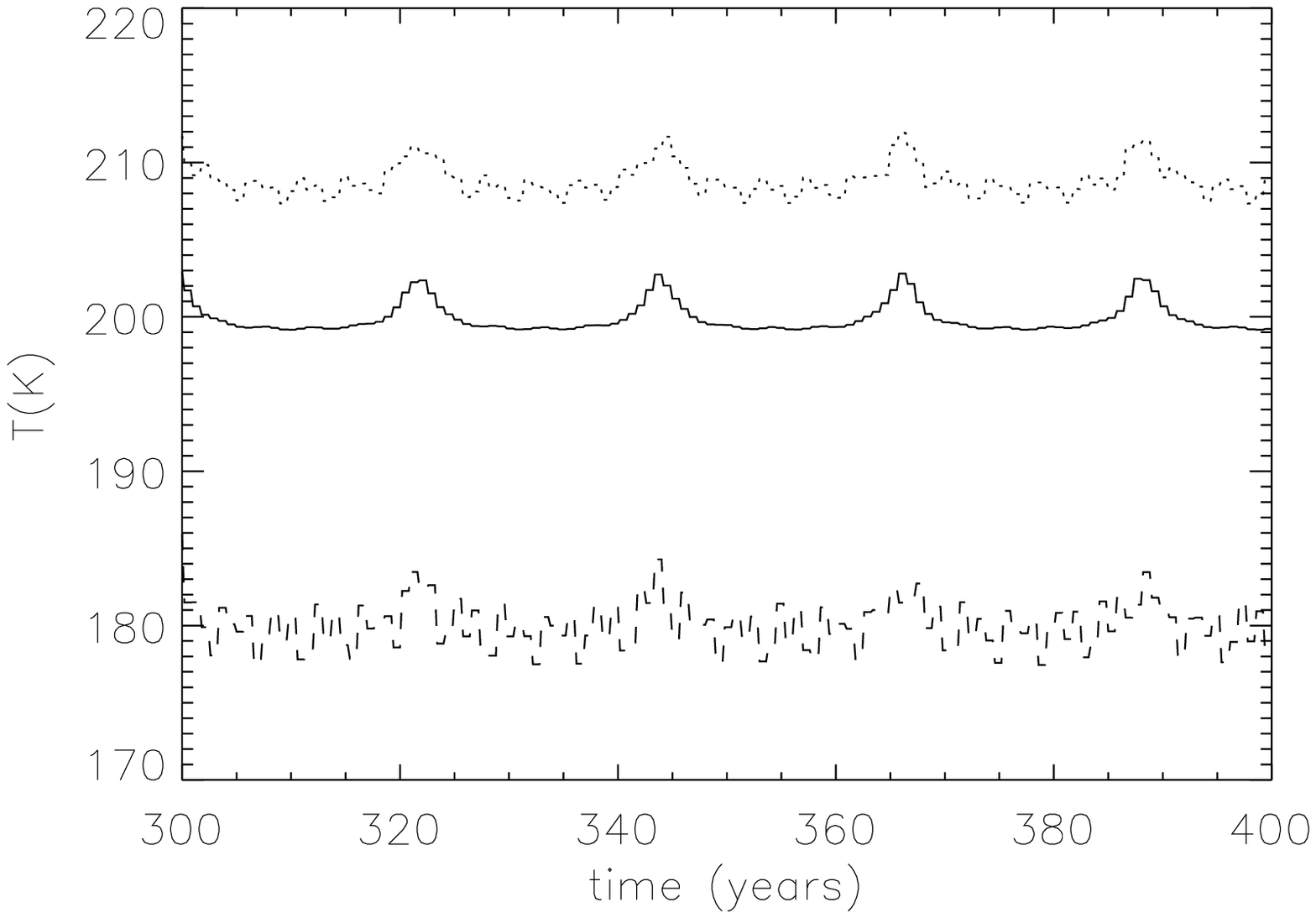} &
\includegraphics[scale=0.45]{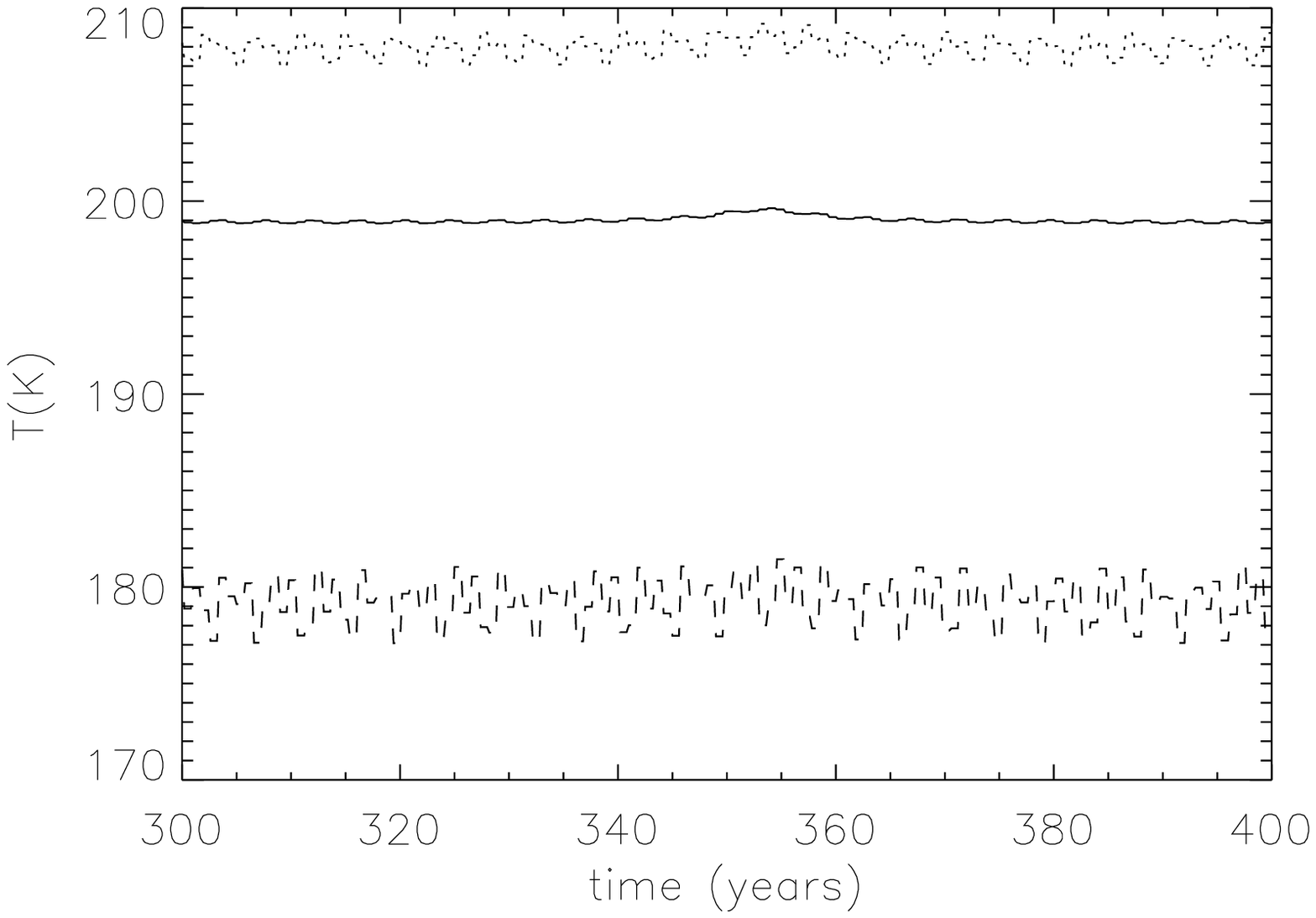}
\end{array}$
\caption{Comparing planet temperature as a function of time as the semi major
  axis of \alphacen A is changed (for a planet with $a_p=0.9$,
  $e=0$).  Top left shows $a_2=5AU$, top right shows $a_2=6AU$, bottom
  left shows $a_2=10 AU$, and bottom right shows the true \alphacen A
  semi major axis of 23.23 AU.  The solid line indicates the global mean temperature, the
  dashed line the minimum temperature, and the dotted line is the
  maximum temperature. \label{fig:compareTbin}}
\end{center}
\end{figure*}

\noindent We have focused specifically on the \alphacen system in this
paper.  It is instructive to investigate other putative binary
systems, to investigate the varying strength of binary perturbations.
As the parameter space of binary star systems is quite large, we
restrict ourselves to varying the orbital parameters of \alphacen A,
$(a_2,e_2)$, and assume the entire system is coplanar.  Figure
\ref{fig:compareTbin} shows how a planet's minimum, maximum and global
mean temperature varies as $a_2$ is decreased (with $e_2$ held at
0.5179).  The planet parameters are fixed at $a_p=0.9$, $e_p=0$.  

Taking the true orbital parameters of \alphacen A (bottom right of
figure), we can see that this is indeed a snowball planet, with the
maximum temperature never exceeding 210 K.  A slight perturbation in
the mean can be seen at $t\sim 350$ years due to passage of \alphacen
A through periastron.  Decreasing $a_2$ also decreases the orbital
period; this can be seen in the other panels of Figure
\ref{fig:compareTbin}, with the minimum $a_2=5$ au (top left).  More
periastron passages induce more temperature fluctuations, but these
are unable to melt the planet from its snowball state, with the
exception of $a_2=5$ au (which produces mean temperatures close to
terrestrial values).  This climate cycle is stable over the entire
simulation time.  In this configuration, the insolation
  produced by \alphacen A constitutes nearly a third of the total
  insolation by both stars at periastron, and around ten percent of
  the total insolation at apastron, so it is not entirely surprising
  that the planet can escape the snowball state.

Whether the outer HZ is extended or not depends quite sensitively on
$a_2$.  We can see that $a_2=6$ au (top right) fails to melt the
planet, despite producing temperature fluctuations of order 10 K.
The insolation due to \alphacen A is still approximately a
  third of the total energy budget at periastron, but the reduced
  frequency of periastron passages due to orbital distance means the
  time-averaged insolation is lower.  This very small change in the
  magnitude and frequency of the perturbation induced by \alphacen A is sufficient
  to completely alter the state of the planet in orbit around
  \alphacen B from habitable to snowball.

In the habitable case shown here, the periastron of \alphacen A is
approximately 2.4 au.  Can we produce another habitable planet by
maintaining this periastron and \alphacen A's true semi major axis,
$a_2=23.23$ au? This would correspond (roughly) to an eccentricity of
$e_2=0.9$.  As the periastron passage is now quite rapid, we should
also expect that the climate model will become much more sensitive to the phase
between the planet's orbit and \alphacen A's.  We investigate the
dependence on phase by running the same simulation with different
initial planetary orbital longitudes ($\phi_p=0\degrees,
90\degrees,180\degrees$). 

\begin{figure*}
\begin{center}$
\begin{array}{c}
\includegraphics[scale = 0.45]{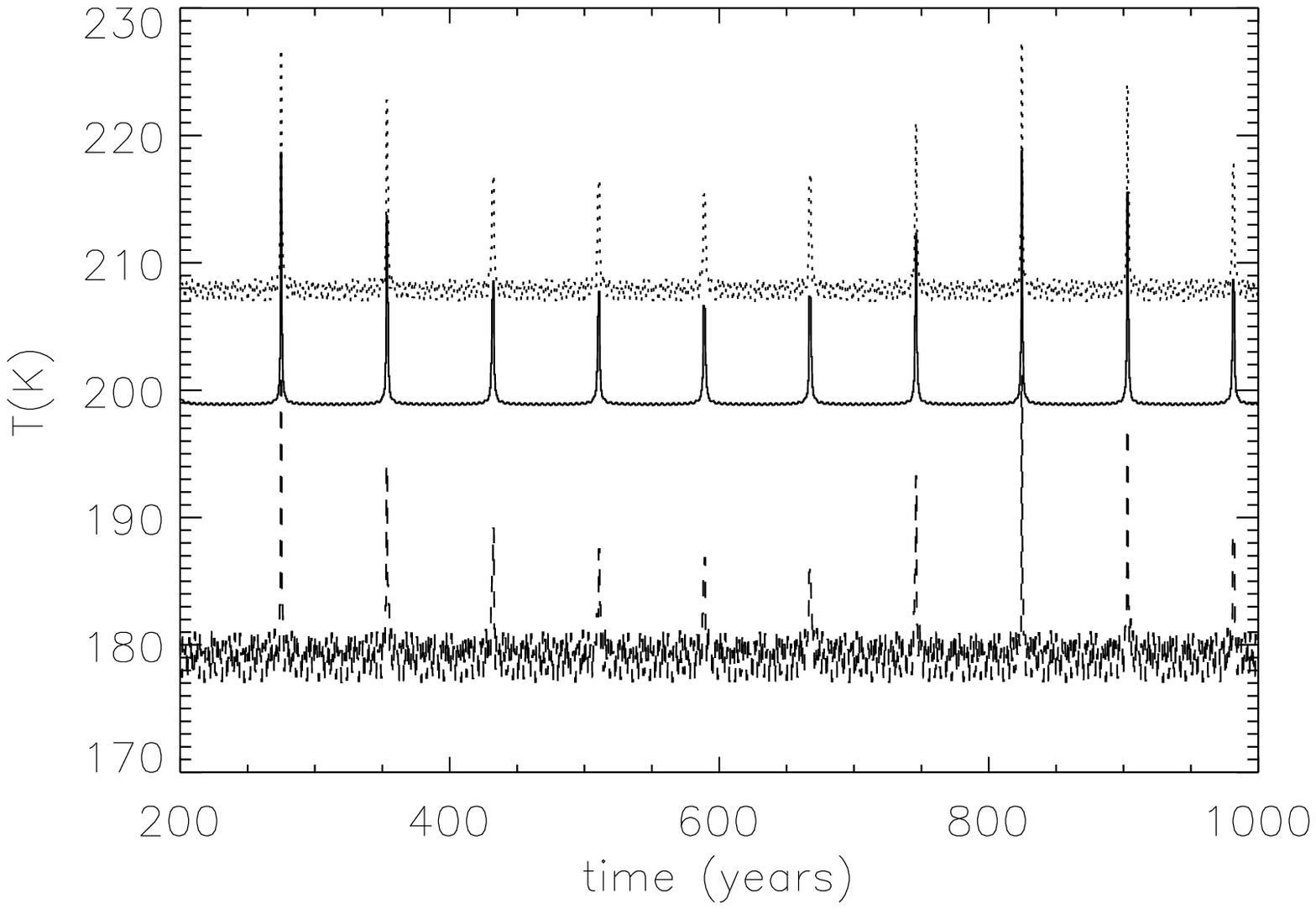} \\
\includegraphics[scale = 0.45]{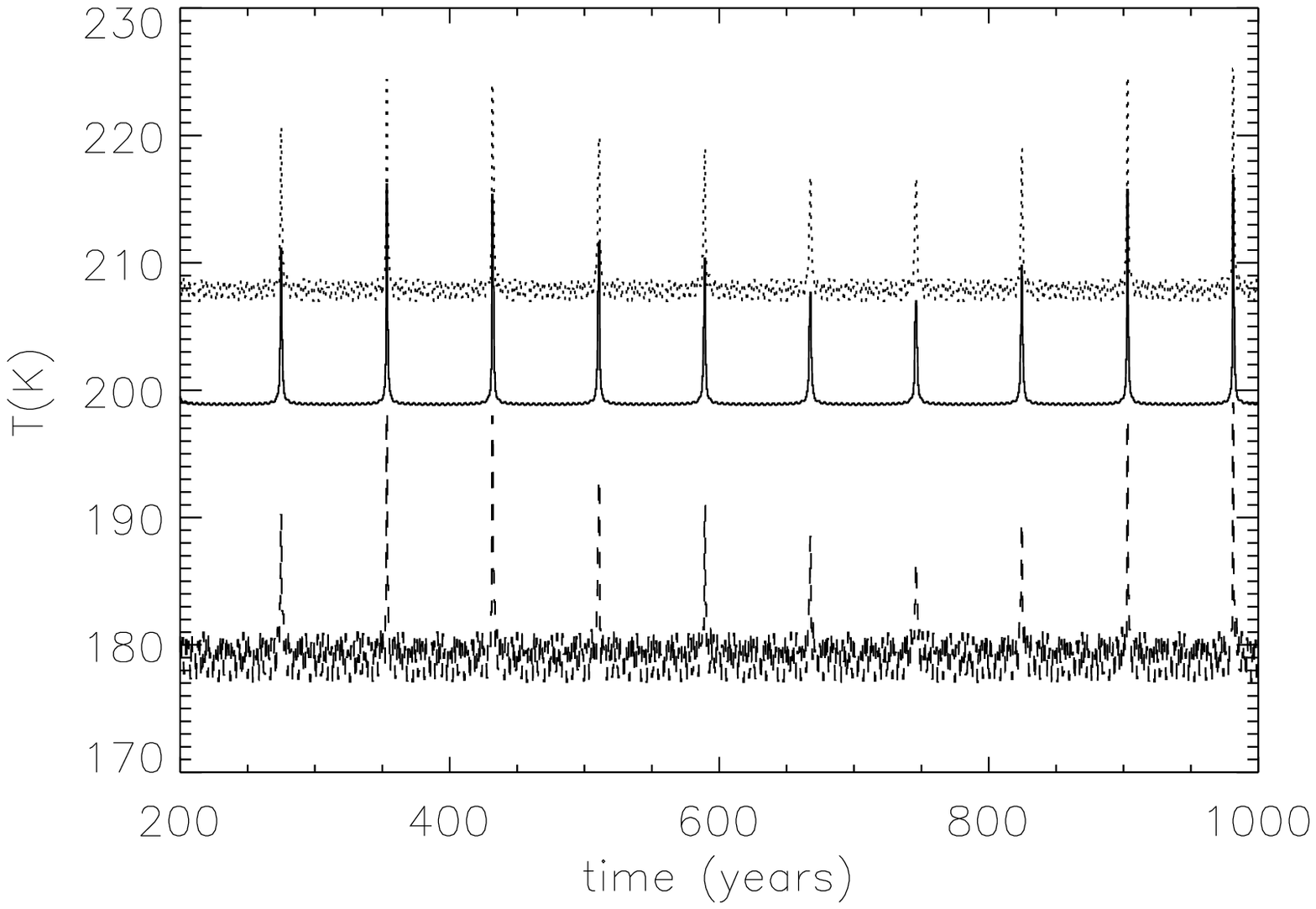} \\
\includegraphics[scale=0.45]{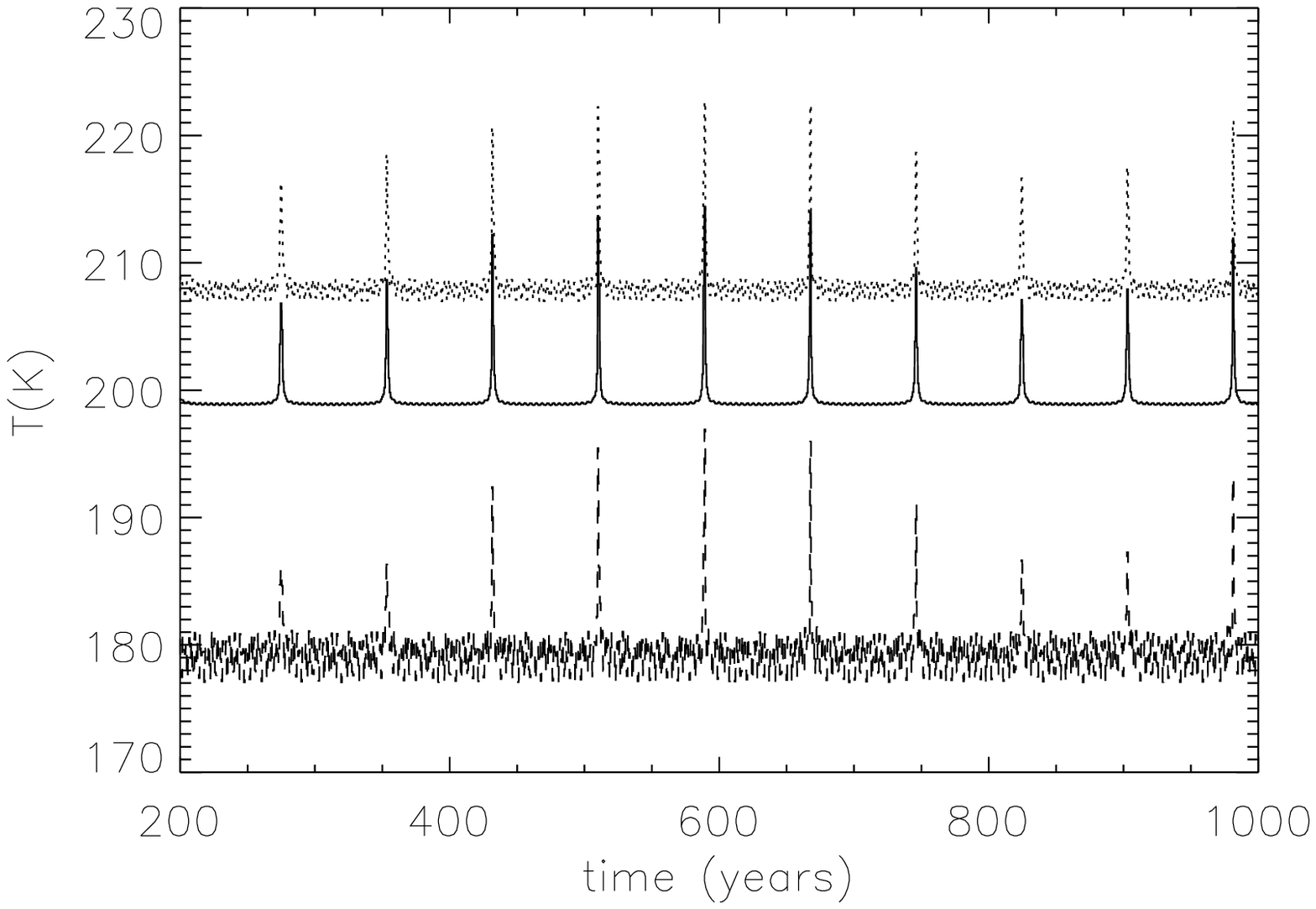} \\
\end{array}$
\caption{Comparing planet temperature as a function of time with the
  eccentricity of \alphacen A, boosted to $e_2=0.9$ (for a planet with
  $a_p=0.9$, $e=0$).  Top shows the starting orbital longitude of the planet
  $\phi=0\degrees$, middle shows $\phi=90\degrees$, and bottom
  shows $\phi=180\degrees$. \label{fig:compareTes09}}
\end{center}
\end{figure*}

The results can be seen in Figure \ref{fig:compareTes09}.  We can see
that the planet remains in a snowball state, but experiences
temperature fluctuations as large as 20 K.  The strength of the
fluctuations clearly varies with the relative position of planet and
secondary as periastron is reached - each of the three plots shows the
same behaviour, but with a phase shift caused by the planet's shifted
starting position.  The perturbation in insolation due to
  \alphacen A retains its maximum value at insolation of around 30\%
  of the total, but the orbital period remains too long for these
  close passages to be sufficiently frequent to melt the planet.

\subsection{Dependence on Ocean Fraction}

\begin{figure}
\begin{center}
\includegraphics[scale = 0.5]{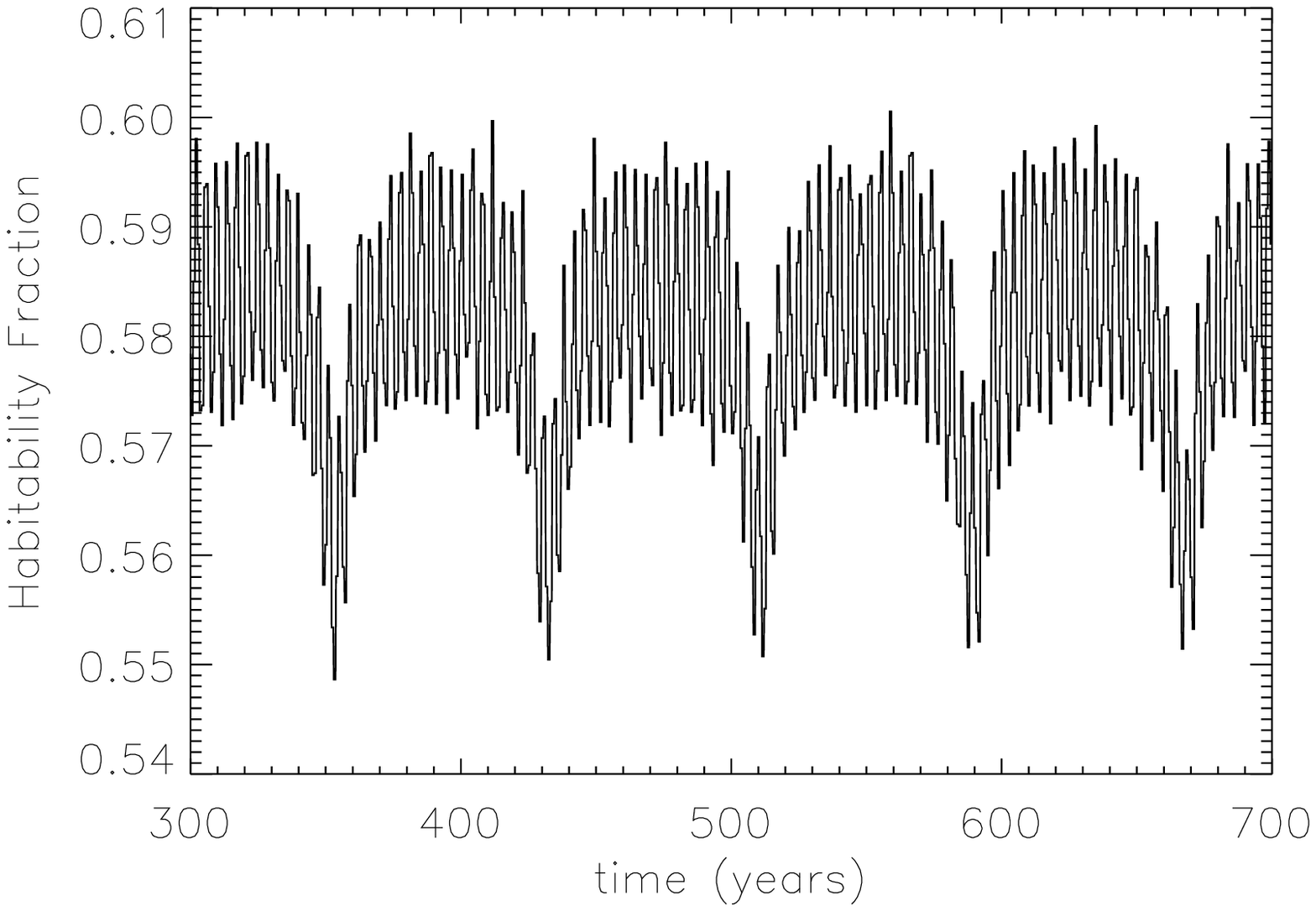}
\caption{Habitability fraction as a function of time for
  the case where $a_p=0.675$, $e_p=0.0$, and the fraction of planet
  surface which is ocean, $f_{ocean}=0.1$. \label{fig:habfrac_dry}}
\end{center}
\end{figure}

\noindent We have assumed that the planets have the same surface ocean
cover as Earth.  Planets with lower ocean fractions are subject to
shorter thermal relaxation timescales, and therefore will experience
greater seasonal variations  \citep{Spiegel_et_al_08}.  It might
therefore be reasonable to assume that drier planets will be more
sensitive to binary fluctuations than wetter planets.  This indeed
appears to be the case, as is shown in Figure \ref{fig:habfrac_dry}. We
display habitability fractions for the same planetary parameters as
Figure \ref{fig:habfrac_bintrans} with $f_{ocean}$ now 0.1.  The
seasonal variations are substantially higher, with the fluctuations
due to \alphacen A also slightly increased (from 0.025 to about 0.03).

\subsection{Dependence on Obliquity}

\noindent The influence of \alphacen A adds a second seasonal
variation to planets around \alphacen B, with the polar declination
varying in a non-trivial fashion over the period of the binary's
orbit.  How will binary oscillations alter under a change in
obliquity? We have used the terrestrial value of $23.5\degrees$ until
this point, but there is no reason to discount other values.  In
our Solar System, Mars' obliquity appears to have varied significantly
between 0$\degrees$ and 60$\degrees$ \citep{Laskar1993}.  In the case of
Earth, the Moon has played an important role in stabilising obliquity
fluctuations \citep{NerondeSurgy1997}, but this is not necessarily a
general result for terrestrial planets with relatively large moons.

The influence of a binary can help lock the planet's obliquity
into a so-called Cassini state \citep{Correia2011}.  While these are
generally low obliquity states, high obliquity states can also occur
\citep{Dobrovolskis2009}.  Even without binary influence, numerical
simulations indicate that the distribution of terrestrial planet
obliquity is isotropic, and therefore primordial obliquities may be
large \citep{Kokubo2007,Miguel2010}.

\citet{Spiegel2009} show that high obliquity planets experience
stronger seasonal variations, and can move far from global radiative
balance even in circular orbits.  While they are not necessarily more
prone to snowball transitions, we may expect them to be more sensitive
to binary fluctuations. 

\begin{figure*}
\begin{center}$
\begin{array}{cc}
\includegraphics[scale = 0.45]{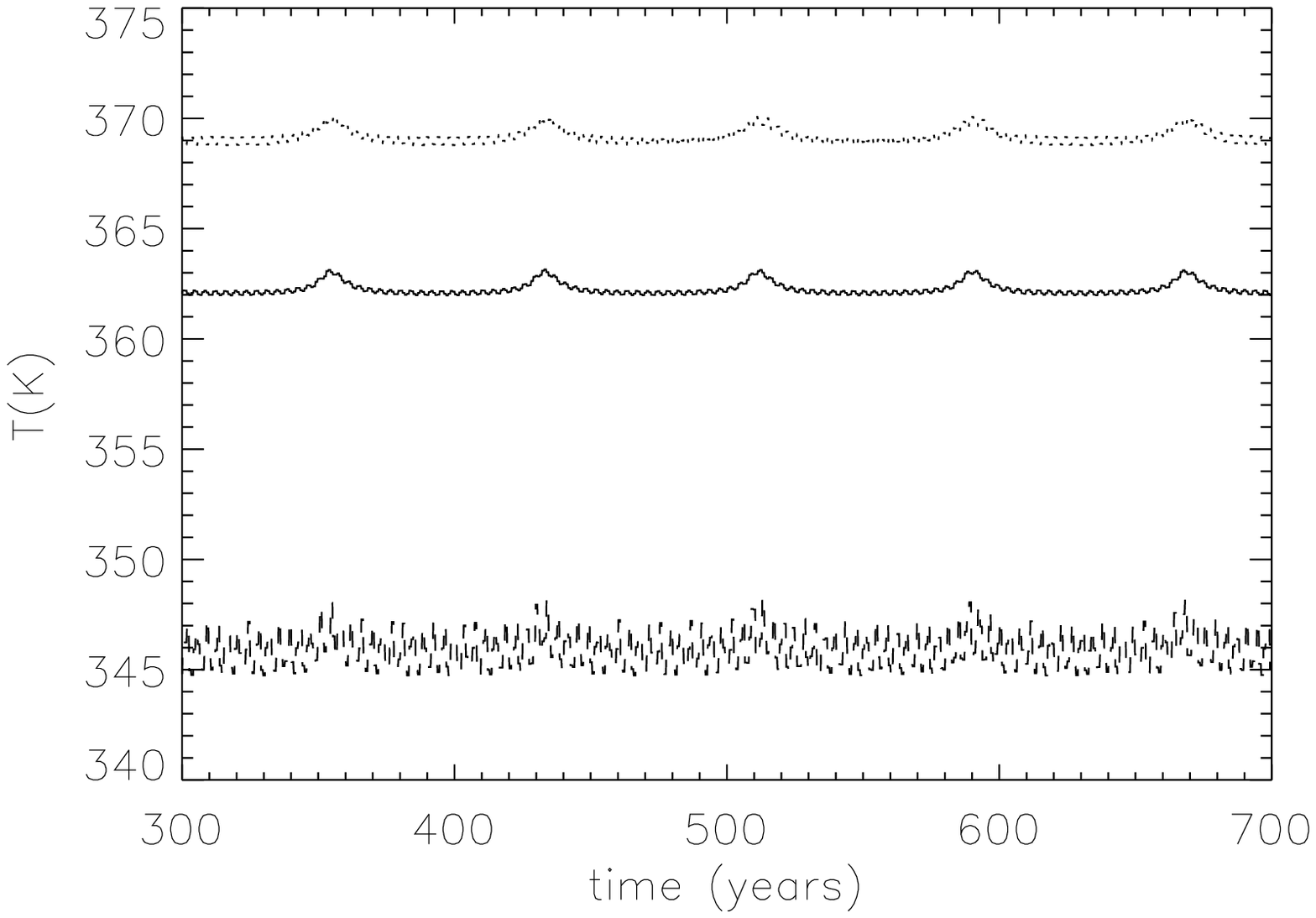} &
\includegraphics[scale = 0.45]{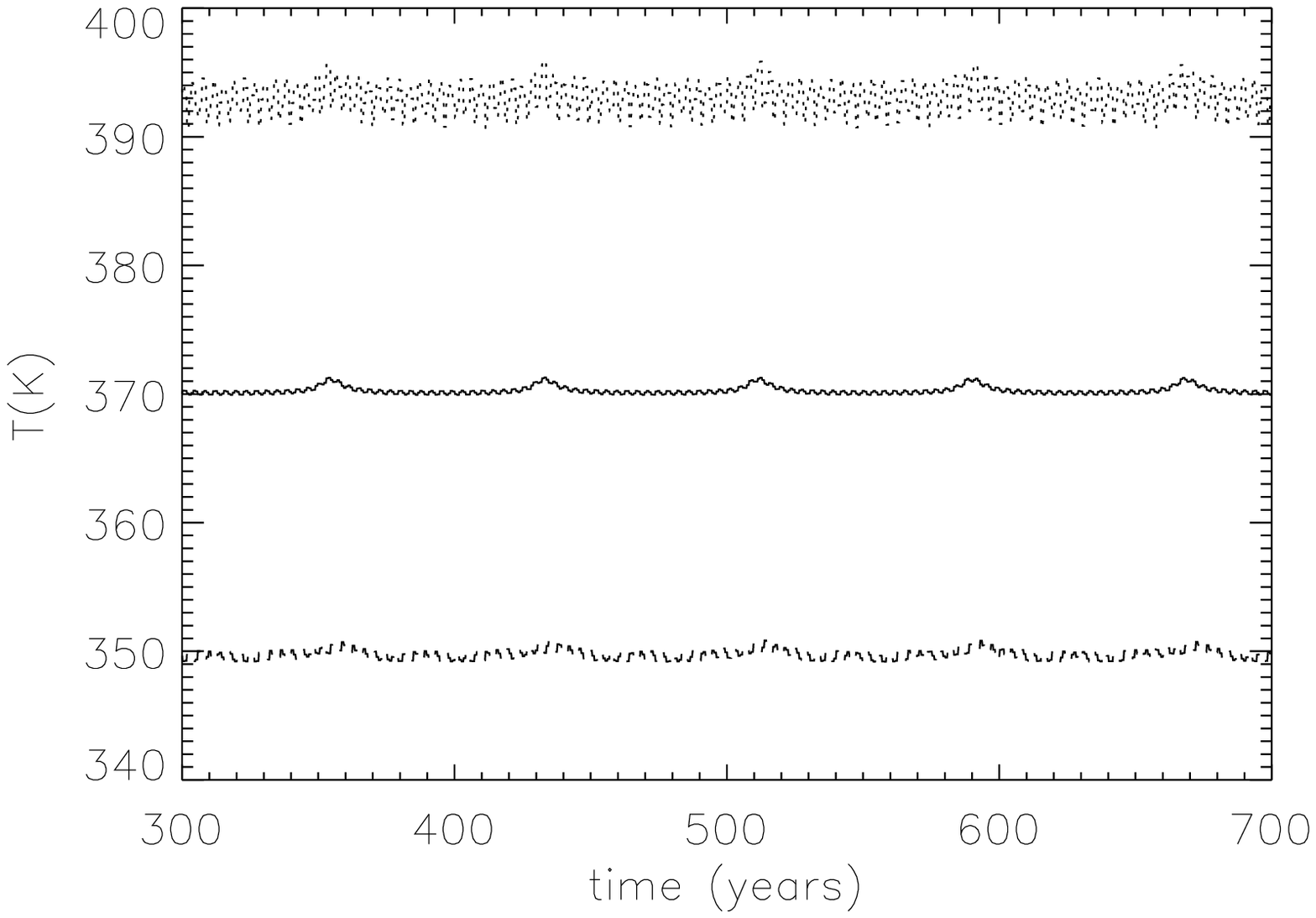} \\
\end{array}$
\caption{Comparing planet temperature as a function of time as the
  obliquity of the planet is changed (for a planet with $a_p=0.675$,
  $e=0$).  The left hand plot shows a planet with with obliquity
  increased to 45$\degrees$, and the right hand plot shows a planet
  with obliquity increased to 90$\degrees$.  The solid line indicates
  the global mean temperature, the dashed line the minimum
  temperature, and the dotted line is the maximum temperature. \label{fig:spin}}
\end{center}
\end{figure*}

We investigate two other values of obliquity for the case where
$a_p=0.675$ au, $e_p=0$, increasing the obliquity to 45$\degrees$ and
$90\degrees$ (Figure \ref{fig:spin}).  The amplitude of the
temperature oscillations are unaffected, but the temperatures
themselves change significantly.  Indeed, in the case of 45$\degrees$
spin, the planet requires re-classification from a binary transient to
a habitable planet.  This change (in line with the results of
\citealt{Spiegel2009} for single-star systems) underlines the
difficulty of classifying planets as habitable or otherwise, as the
parameter space for habitability is non-trivial and high in dimension.

\subsection{Dependence on Rotation Rate}

\noindent Other than convenience, there are no real reasons to assume
that planets around \alphacen B possess the same diurnal period as
Earth.   

\begin{figure*}
\begin{center}$
\begin{array}{c}
\includegraphics[scale = 0.45]{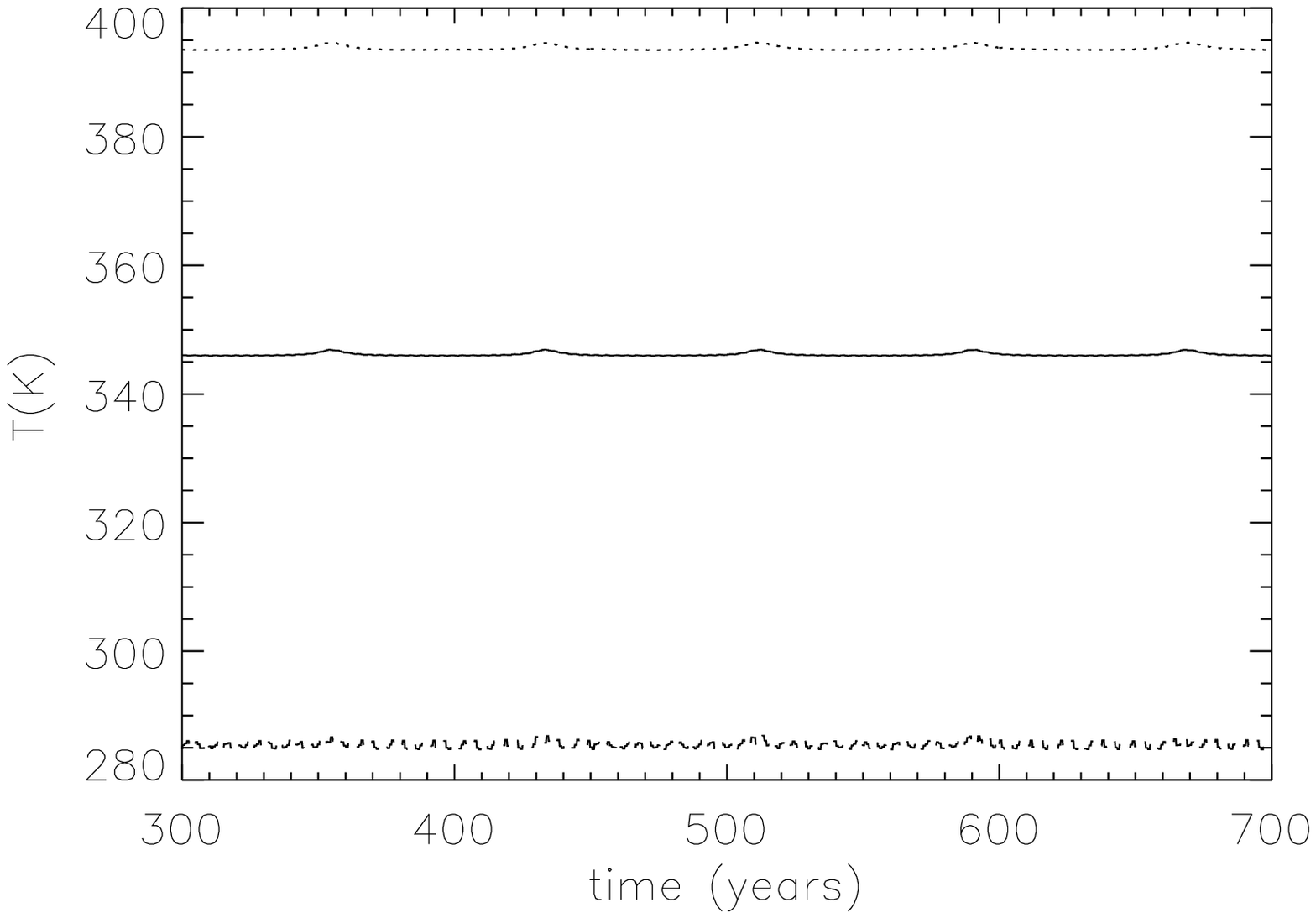} \\
\includegraphics[scale = 0.45]{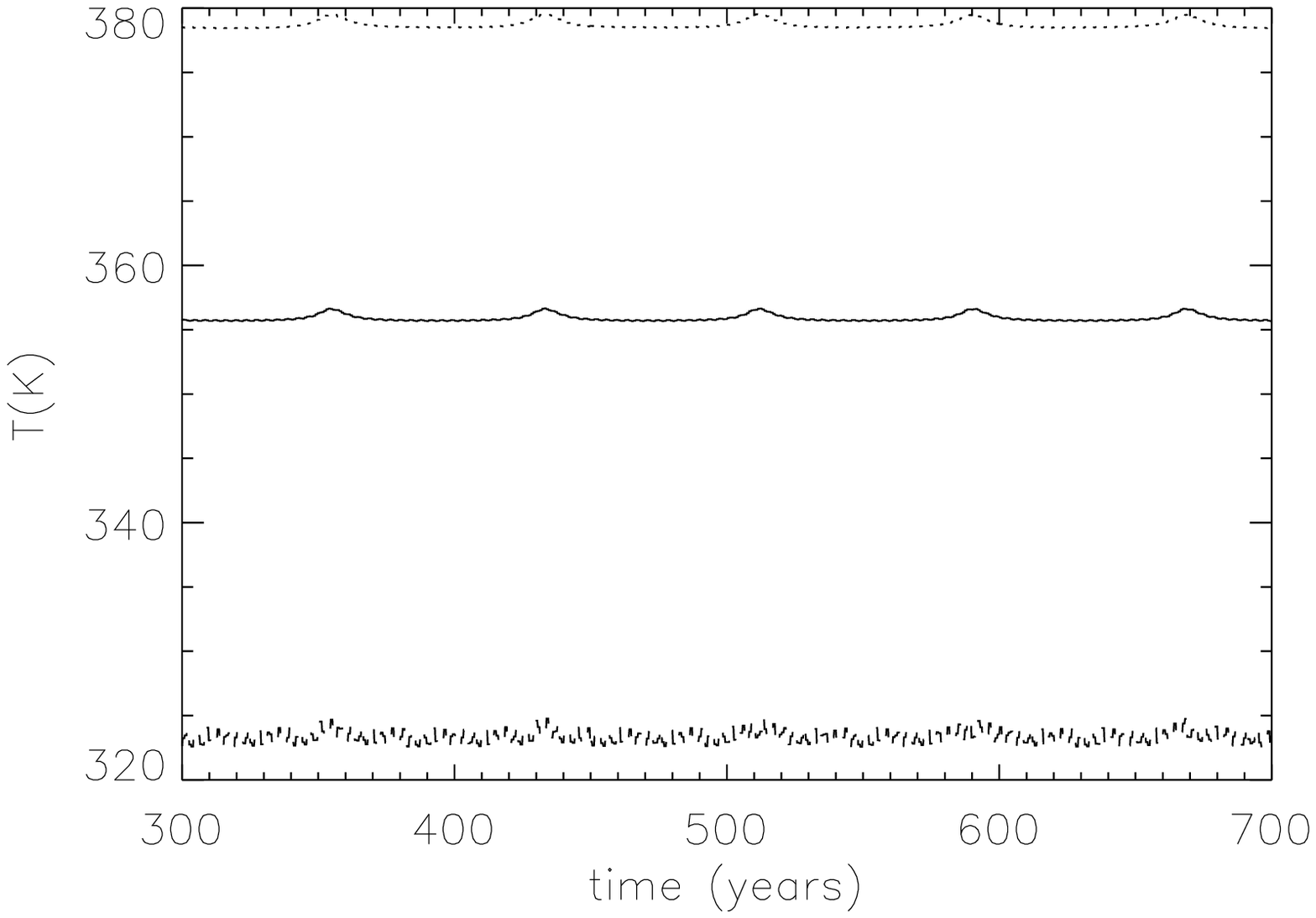} \\
\includegraphics[scale = 0.45]{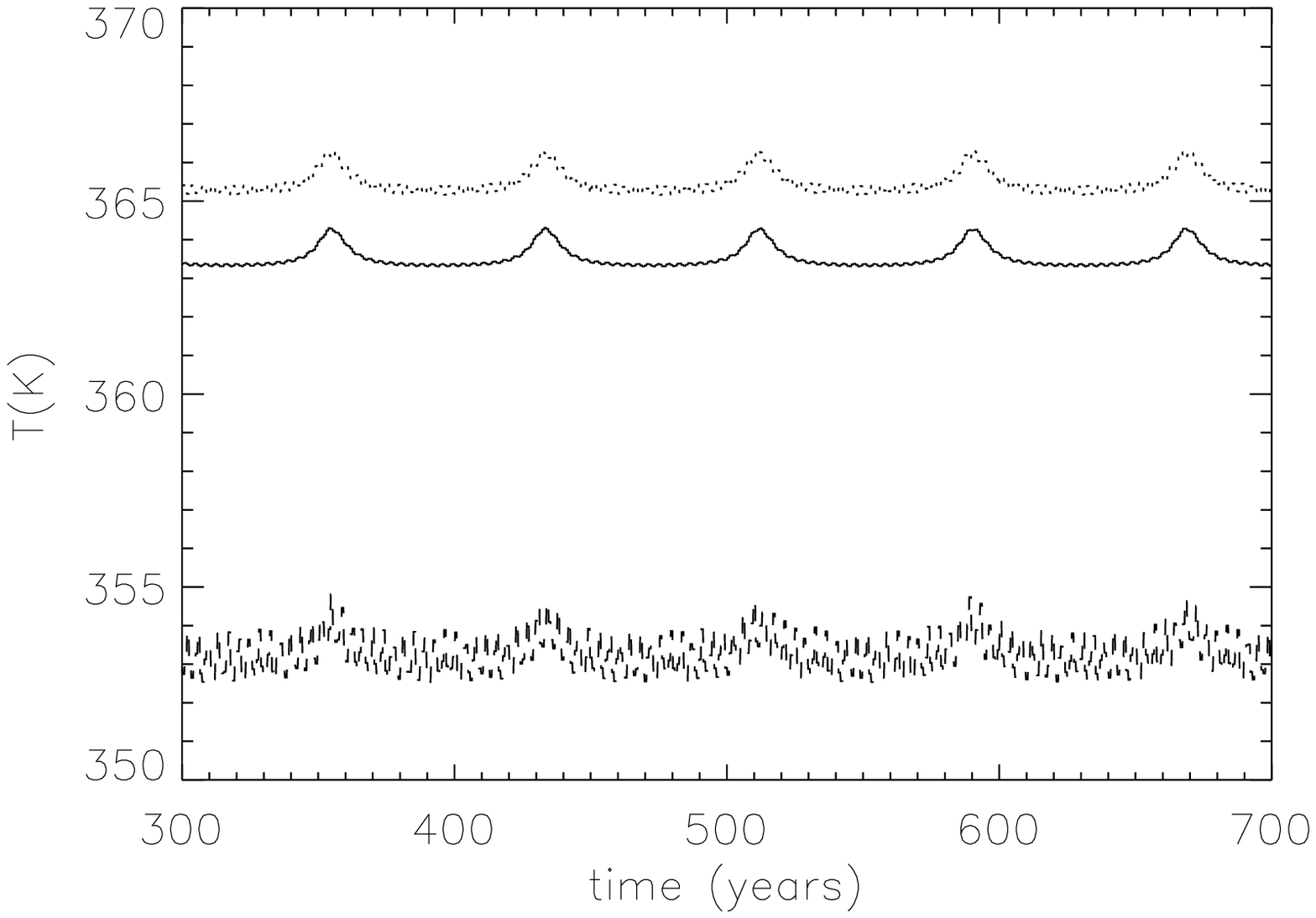} \\
\end{array}$
\caption{Comparing planet temperature as a function of time as the diurnal
  period of the planet is changed (for a planet with $a_p=0.675$,
  $e=0$).  The left hand plot shows a planet with diurnal period equal
  to one day, and the right hand plot shows a planet with diurnal
  period equal to a third of a day.  The solid line indicates the
  global mean temperature, the dashed line the minimum temperature,
  and the dotted line is the maximum temperature. \label{fig:compareTrot}}
\end{center}
\end{figure*}

We investigate planets rotating with 8, 24 and 72 hour periods
(with $a_p=0.675$, $e_p=0$).  Figure \ref{fig:compareTrot} shows the
resulting minimum, maximum and mean temperatures as a function of time
for all three cases.  The temperature fluctuation induced by \alphacen A
maintains the same amplitude regardless of rotation rate.  

Reducing the rotation period reduces heat transport, causing the radiant energy
deposited by the primary to be diffused to a smaller range of
latitudes, increasing the temperature gradient in the fast rotating
case (top panel of Figure \ref{fig:compareTrot}).  The slower rotating
case shows the more efficient latitudinal transport distributing
insolation such that the temperature range from pole to pole is less
than 20K (bottom panel).  

\subsection{Limitations of the Model}

%\noindent Limitations of the model.  Expected amplitude variations of the
%planet's orbit due to secondary perturbations? Lack of carbon-silicate
%cycle - could this extend habitable zone? Damp binary oscillations?

\noindent Finally, we should note the limitations of the above
analyses.  As with the work of \citet{Spiegel_et_al_08}, the 1D LEBM
can only capture a restricted range of thermal timescales.  While it
is sensitive to seasonal forcing and (as a result) orbital timescales,
the model does not exhibit the features of much slower climate
processes, which possess timescales many times the orbital period of
the \alphacen system.  For example, oceanic circulation is not included in this
model, which can determine longer-term climate variation.

More complex additions, such as clouds and a carbonate-silicate cycle
(e.g. \citealt{Williams1997a}) would provide a further regulating
effect on planet temperatures, potentially extending the outer edge of
the habitable zone \citep{Kasting_et_al_93} on long enough timescales.
In the particular case of binary systems, we might expect that the
oscillations induced by the presence of a companion would be damped,
if the silicate cycle can respond sufficiently quickly during
periastron passage.  The timescale on which the silicate cycle
  can be expected to respond is sensitive to the specific properties
  of the planet, i.e. its chemistry, geology and ocean circulation
  systems.  For the Earth, it has been estimated that the
  equilibration timescale for the carbonate-silicate cycle is around
  0.5 Myr \citep{Williams1997a}.  A more rapid means of $C0_2$ release
  may be through warming the oceans, as their capacity to hold $C0_2$
  decreases with increasing temperature. The relevant timescale would
  instead be ocean circulation timescales, which are of the order
  $10^3$ years.  Again, these timescales are representative for the
  Earth only, and will not apply in general to terrestrial
  exoplanets.

Perhaps the most obvious limitation of these models is their
dimension.  1D modelling prevents discussion of longitudinal climate
variations, and forces us to consider diurnally averaged insolation.
This prevents simulation of planets with slow rotation rates relative
to their orbital motion (such as Venus, see \citealt{Parish2011}).
The contrast between land and ocean is also lost in 1D, and
latitudinal variations that occur as a result are not accounted for.
Despite this, 1D LEBMs still capture much of the relevant physics,
capable of reproducing fiducial Earths with temperature profiles very
similar to real data \citep{Spiegel_et_al_08}.  The inner edge of the
habitable zone is less well-defined than the outer edge - atmospheric
changes could allow liquid water above 373 K, and the runaway
greenhouse effect may become important at temperatures nearer 350 K
(\citealt{Spiegel_et_al_08} and references within).  In any case, the
outer edge is likely to be more interesting from an astrobiological
standpoint, as current and future instrumentation will be more capable
of probe spectral features of planets at larger semi-major axes (see
e.g. \citealt{Kaltenegger2010}).
   
We should acknowledge, however, that the addition of a second
insolation source into diffusion approximation-based models such as
this is not entirely understood.  However, the perturbations induced
by secondary insolation are relatively small (typically a few percent
of the primary insolation, except in extreme cases).  We therefore
argue that the models are appropriate for an initial exploratory
investigation.  

 \citet{Thevenin2002} estimate the luminosity of \alphacen B as
 $0.5002 \pm 0.016 L_{\rm \odot}$, which differs from the main
 sequence value used in this work by around 30\%.  The habitable zone
 would be somewhat closer to the star as a result, increasing the
 number of planetary orbits per binary orbit.  While much of the
 qualitative trends in this paper would remain unchanged, it is
 clearly important that future studies of habitability in this system
 use observationally constrained luminosities.

We should also note that we neglect the effect of \alphacen A's
gravitational field on the dynamics of \alphacen B's planets.  While
the planet's orbital inclination is low enough to avoid Kozai
resonances, and the mean motion resonances due to \alphacen A are
absent from the orbits we explore \citep{Michtchenko2009}, we should
still expect short period perturbations to the planet semi-major axis
of order 0.01 au, more than sufficient to produce substantial
Milankovitch cycles \citep{Spiegel2010}.  We have seen that the outer
edge of the HZ is sensitive to perturbations of this size, so future
studies must incorporate these perturbations in more detail.

\section{Conclusions}\label{sec:Conclusions}

\noindent We have investigated the influence of \alphacen A on the
habitable zone of \alphacen B, to test the efficacy of the single-star
approximation in a binary context.  In general, we demonstrate that
the single-star approximation is roughly correct for calculating the
inner and outer boundaries of the habitable zone, but fails to capture
oscillations in the planet's climate that occur as a result of
\alphacen A's passage through periastron.    

For all planetary orbits, the presence of \alphacen A induces
temperature fluctuations of order a few K.  At the habitable zone
boundaries, the fraction of habitable surface on such planets can be
altered by around 3\%.  The strength of these fluctuations can be
increased by reducing the planet's ocean fraction, or increasing its
obliquity.  

It is reasonable to speculate that if life were to exist on planets
around \alphacen B, that they may develop two circadian rhythms (cf
\citealt{Breus1995}) corresponding to both the length of day around
the primary, and the period of the secondary's orbit (approx 70
years).  Altering the available habitat by a few percent may also
influence migration patterns and population evolution.

While we have demonstrated that the temperature fluctuations for
planets around \alphacen B due to \alphacen A are relatively small,
the consequences of a periodic temperature forcing of a few K to long
term climate evolution cannot be fully understood from this work.  To
fully appreciate the impact on (for example) ocean circulation and
carbonate-silicate cycles requires further investigation with more
advanced climate models.

\section*{Acknowledgments}

\noindent All simulations were performed using high performance
computing funded by the Scottish Universities Physics Alliance (SUPA).
DF gratefully acknowledges support from STFC grant ST/H002380/1.  The
author would like to thank Caleb Scharf for useful discussions, and
the referee, David Spiegel, whose comments and insights greatly
improved this manuscript.

\bibliographystyle{mn2e} % (must include a bibliography style)
\bibliography{alpha_centauri}

\appendix

\label{lastpage}

\end{document}